\documentclass[11pt,a4paper]{article}
\pdfoutput=1
\usepackage{jstyle}
\usepackage{amsthm,bm,mathrsfs}
\usepackage{youngtab,simplewick}

\author{\quad Karapet Mkrtchyan\footnote{At Scuola Normale Superiore di Pisa (Italy) from September 2019. Email: karapet.mkrtchyan@sns.it}}

\affiliation{Max Planck Institute for Gravitational Physics (Albert Einstein Institute)\\ Am M\"uhlenberg 1, 14476 Potsdam, Germany}

\emailAdd{karapet.mkrtchyan@aei.mpg.de}

\title{\centering 
On Covariant Actions for Chiral $p-$Forms}

\abstract{
We construct a Lorentz and generally covariant, polynomial action for free chiral $p-$forms, classically equivalent to the Pasti-Sorokin-Tonin (PST) formulation. The minimal set up requires introducing an auxiliary $p-$form on top of the physical gauge $p-$form and the PST scalar. The action enjoys multiple duality symmetries, including those that exchange the roles of physical and auxiliary $p-$form fields. Actions of the same type are available for duality-symmetric formulations, which is demonstrated on the example of the electromagnetic field in four dimensions. There, the degrees of freedom of a single Maxwell field are described employing four distinct vector gauge fields and a scalar field.
\\
\vskip0.5cm

\rightline{\it To my mother}
}

\begin{document}

\maketitle

\section{Introduction}

The progress in our understanding of a given theoretical framework often happens when we tackle the specific ``marginal'' examples where the framework fails in its standard form. In Field Theory, a problem of this kind is related to chiral $p-$forms and their interactions, also related to the problem of manifesting (electric-magnetic) duality in familiar field theories.
 
Duality-symmetric formulations and chiral $p-$forms have been studied to great extent in the past forty years (see \cite{Zwanziger:1970hk,Deser:1976iy,Marcus:1982yu,Siegel:1983es,Kavalov:1986ki,Floreanini:1987as,Henneaux:1988gg,Harada:1989qp,Tseytlin:1990nb,McClain:1990sx,Wotzasek:1990zr,Tseytlin:1990va,Schwarz:1993vs,Khoudeir:1994zw,Pasti:1995ii,Pasti:1995tn,Pasti:1995us,Tseytlin:1996it,Devecchi:1996cp,Pasti:1996vs,Cederwall:1997ab,Maznytsia:1998xw,Pasti:1997mj,Rocek:1997hi,Manvelyan:1999vr,Kuzenko:2000tg,Miao:2000fn,Sorokin:2002rsa,Bunster:2012hm,Afshar:2018apx} for a non-comprehensive list of key references).
It is known that the construction of space-time covariant actions for free chiral p-forms requires the use of auxiliary fields. There are several formulations which deal with this issue. The most economic and efficient one is that of PST \cite{Pasti:1995ii,Pasti:1995tn,Pasti:1996vs}. It uses a single auxiliary scalar field which however enters the action (under derivatives) in a non-polynomial way. 
In this paper we will provide  a simple polynomial action which contains more auxiliary fields and is classically equivalent to that of PST.
It combines manifest Lorentz covariance, finite number of fields, polynomial form, no ghosts and consistence with general covariance. 

Exploring alternative formulations for free field theories may open new possibilities for their interacting extensions. The simplest example to demonstrate this is the familiar scalar field with its $(d-2)$-form gauge field dual (so-called ``notoph'' \cite{Ogievetsky:1967ij}). While for a scalar field one can immediately formulate interacting theories with arbitrary non-derivative interaction potential, the dual formulation via $(d-2)$-form gauge field does not admit non-derivative interactions. Therefore, if we were to start with the $(d-2)-$form, we would not be able to construct, e.g., the massless scalar field theory with $\phi^4$ interaction. The long-standing problem of non-abelian interactions for chiral $p-$forms motivates the search for a suitable formulation of the free theory, which will allow for non-abelian extensions.

Recently, inspired by String Field Theory, Ashoke Sen devised an action \cite{Sen:2015nph,Sen:2019qit} for chiral $p-$forms, that has all of the aforementioned virtues except for general covariance. Even though it contains an extra ghost, the latter decouples from the physical sector, therefore allowing for unitary dynamics of the physical degrees of freedom. The field variable, containing the chiral $p-$form degrees of freedom for the Sen's model is the $(p+1)$-form field strength. 
Possible non-abelian extensions for chiral $p-$forms would likely prefer potentials as basic variables instead, similarly to what happens in Yang-Mills theory. It is indeed only possible to write Born-Infeld type interactions through curvatures, while the minimal couplings of Yang-Mills type are available only in terms of gauge-variant potentials. %We therefore expect that the formulation provided here is a good starting point to investigate interactions of chiral p-forms. 

The action we derive here has all of the properties listed above, and is written through gauge potentials, therefore is a good starting point in the search for an interacting extension that would describe non-abelian chiral $p-$form theories.

%In particular, for 2-forms, there is an expected non-abelian theory underlying $(2,0)$ model \cite{} in six dimensions, that is supposed to describe the world-volume theory of multiple $M5-$branes \cite{}.

The formulation we study here applies also to duality-symmetric theories. We will show it on the example of Maxwell theory in four dimensions in Section \ref{Maxwell4d}.

\section{Polynomial action for free chiral $p-$forms}

We start with a Lagrangian for a chiral $p-$form $\varphi_{\m_1\dots\m_p}$ in $d=2p+2$ dimensions \footnote{For chiral forms, we take $p$ to be even. A similar action will be discussed in the following for duality-symmetric odd-forms which require doubled set of fields. For odd $p$, the chiral fields do not exist in Minkowski background. The regime of validity of our formulation is the same as that of the PST formalism --- it is only available in Minkowski signature. We thank Dmitri Sorokin for correspondence on this matter.}:
\begin{align}
\mathcal{L}=-\frac1{2(p+1)}\, F_{\m_1\dots\m_{p+1}}\,F^{\m_1\dots\m_{p+1}}-\frac1{2(p+1)}\,(\mathcal{F}_{\m_1\dots\m_{p+1}}-(p+1)\,c_{[\m_1}\,R_{\m_2\dots\m_{p+1}]})\times\nonumber\\\times\,(\mathcal{F}^{\m_1\dots\m_{p+1}}-(p+1)\,c^{[\m_1}\,R^{\m_2\dots\m_{p+1}]})+G^{\m\n}\partial_{[\m}c_{\n]}\,,\qquad\label{Chiralp}
\end{align}
where
\begin{align}
F_{\m_1\dots\m_{p+1}}=(p+1)\,\partial_{[\m_1}\varphi_{\m_2\dots\m_{p+1}]}\,,\\
\mathcal{F}_{\m_1\dots\m_{p+1}}=F_{\m_1\dots\m_{p+1}}+\frac{1}{(p+1)!}\e_{\m_1\dots\m_{p+1}\n_1\dots\n_{p+1}}\,F^{\n_1\dots\n_{p+1}}\,,
\end{align}
while $c_\m$, $R_{\m_1\dots\m_{p}}$ and $G^{\m\n}$ are auxiliary fields with fully antisymmetric set of Lorentz indices. Even though the Lagrangian given above is not quadratic in fields, it is quadratic in the physical gauge potential $\varphi_{\m_1\dots\m_p}$ and can be shown to describe exactly a single chiral degree of freedom, in Minkowski space of $2p+2$ dimensions for even\footnote{For even $p$ the chiral (self-dual) fields are non-trivial in Minkowski spaces and for odd $p$ --- in Euclidean spaces. We will not discuss other possible signatures except for Minkowski here. We choose mostly plus signature convention for the metric.} $p$.
%The algebra of gauge symmetries of the Lagrangian \eqref{Chiralp} will be studied in full details elsewhere. 

\subsection{Equivalence to PST}

We will first show that \eqref{Chiralp} describes a free chiral $p-$form in $d=2p+2$ dimensions. For that, we can integrate out the auxiliary field $R_{\m_1\dots \m_{p+1}}$ and end up with an equivalent non-polynomial Lagrangian of the PST form \cite{Pasti:1996vs}. In order to see that, we solve the algebraic equation of motion for the field $R_{\m_1\dots\m_p}$,
\begin{align}
\mathcal{F}_{\m_1\dots\m_{p+1}}\,c^{\m_1}-c^2\,R_{\m_2\dots\m_{p+1}}+(-1)^{p+1}\,p\,c_{[\m_2}\,R_{\m_3\dots\m_{p+1}]\m_1}\,c^{\m_1}=0\,.
\end{align}
Due to an (algebraic!) gauge symmetry of the Lagrangian \eqref{Chiralp}, given via 
\begin{equation}
\d R_{\m_1\dots\m_p}=c_{[\m_1}\l_{\m_2\dots\m_p]}\,,\label{AGT}
\end{equation}
the solution is fixed only up to an arbitrary $(p-1)-$form field $\l_{\m_1\dots\m_{p-1}}$:
\begin{align}
R_{\m_1\dots\m_p}=\frac{1}{c^2}\mathcal{F}_{\n\m_1\dots\m_p}\,c^\n+c_{[\m_1}\l_{\m_2\dots\m_p]}\,.\label{SolR}
\end{align}
A simple way to solve for $R_{\m_1\dots\m_p}$ is to choose a gauge $R_{\m_1\dots\m_p}\,c^{\m_1}=0$, then the solution is \eqref{SolR} without the last term, which reflects the gauge freedom.
We can now plug back \eqref{SolR} to \eqref{Chiralp} and arrive at the following Lagrangian:
\begin{align}
\mathcal{L}=-\frac1{2(p+1)}\, F_{\m_1\dots\m_{p+1}}\,F^{\m_1\dots\m_{p+1}}+\frac1{2\, c^2}\,\mathcal{F}_{\m_1\dots\m_p\n}\,c^{\n}\,\mathcal{F}^{\m_1\dots\m_p\r}\,c_{\r}+G^{\m\n}\partial_{[\m}c_{\n]}\,,\label{ChiralPST}
\end{align}
It is a trivial exercise to show the equivalence of the Lagrangian \eqref{ChiralPST} with that of \cite{Pasti:1996vs}.
%One advantage of the polynomial form \eqref{Chiralp} is that there is no unwanted divergence of the action in the $c^2\rightarrow 0$ limit. 
It is straightforward to integrate out the Lagrange multiplier field $G^{\m\n}$ in both \eqref{Chiralp} and \eqref{ChiralPST}, solving the zero-curvature equation for $c_\m$ as $c_\m=\partial_\m\,a$ and plugging back into the action. The minimal polynomial form thus contains only two auxiliary fields on top of the physical field $\varphi$ --- the $p-$form $R_{\m_1\dots\m_p}(x)$ and the scalar $a(x)$:
\begin{eqnarray}
\mathcal{L}=-\frac1{2(p+1)}\, F_{\m_1\dots\m_{p+1}}\,F^{\m_1\dots\m_{p+1}}-\frac1{2(p+1)}\,(\mathcal{F}_{\m_1\dots\m_{p+1}}-(p+1)\,\partial_{[\m_1} a\,R_{\m_2\dots\m_{p+1}]})\times\nonumber\\\times\,(\mathcal{F}^{\m_1\dots\m_{p+1}}-(p+1)\,\partial^{[\m_1} a\,R^{\m_2\dots\m_{p+1}]})\,,\qquad\label{ChiralpMin}
\end{eqnarray}
Integrating out $G^{\m\n}$ in \eqref{ChiralPST}, we get the familiar PST form. This step involves solving a differential equation: the two actions are not guaranteed to admit equivalent interactions.

Note, that $\cF^2=\cF\wedge \cF=0$. In fact, the simplest polynomial extension for the PST Lagrangian \eqref{ChiralPST} would be the following one:
\begin{align}
\mathcal{L}=-\frac1{2(p+1)}\, F_{\m_1\dots\m_{p+1}}\,F^{\m_1\dots\m_{p+1}}-\frac1{2}\,\mathcal{F}_{\m_1\dots\m_{p+1}}\,c^{\m_1}\,R^{\m_2\dots\m_{p+1}}+\frac14\,c^2\,R^{\m_1\dots\m_p}\,R_{\m_1\dots\m_p}+G^{\m\n}\partial_{[\m}c_{\n]}\,,\label{PSTpoly}
\end{align}
which differs from \eqref{Chiralp} by a  term $(R_{\m_1\dots\m_p}\,c^{\m_p})^2$, essential for the gauge symmetry \eqref{AGT}. Addition of that term promotes the second class constraint $R_{\m_1\dots\m_p}\,c^{\m_p}=0$ to first class: while in \eqref{PSTpoly} it is a consequence of the equations of motion (similar to divergence-free condition $\partial^\m A_\m=0$ of the vector field in Proca theory), in \eqref{Chiralp} it is a gauge choice (analogous to the Lorentz gauge condition $\partial^\m A_\m=0$ in Maxwell theory).

\subsection{Manifestation of two abelian gauge symmetries}

The Lagrangian \eqref{ChiralpMin} can be recast in the following form:
%\begin{eqnarray}
%\mathcal{L}=-\frac1{2(p+1)}\, F_{\m_1\dots\m_{p+1}}\,F^{\m_1\dots\m_{p+1}}+F_{\m_1\dots\m_{p+1}}\partial^{[\m_1}a\,R^{\m_2\dots\m_{p+1}]}\nn-\frac{(p+1)}{2}\partial_{[\m_1} a\,R_{\m_2\dots\m_{p+1}]}\partial^{[\m_1} a\,R^{\m_2\dots\m_{p+1}]}\nn+\frac{1}{(p+1)!}\e_{\m_1\dots\m_{p+1}\n_1\dots\n_{p+1}}F^{\n_1\dots\n_{p+1}}\partial^{\m_1} a\,R^{\m_2\dots\m_{p+1}}\,,
%\end{eqnarray}
%or, equivalently, as
\begin{eqnarray}
\mathcal{L}=-\frac1{2(p+1)}\, (F_{\m_1\dots\m_{p+1}}-(p+1)\partial_{[\m_1}a\,R_{\m_2\dots\m_{p+1}]})\,(F^{\m_1\dots\m_{p+1}}-(p+1)\partial^{[\m_1}a\,R^{\m_2\dots\m_{p+1}]})\nn+\frac{1}{(p+1)!}\e_{\m_1\dots\m_{p+1}\n_1\dots\n_{p+1}}F^{\n_1\dots\n_{p+1}}\partial^{\m_1} a\,R^{\m_2\dots\m_{p+1}}\,,\qquad\label{ChiralpIntermediate}
\end{eqnarray}
which, after a field redefinition $\varphi_{\m_1\dots\m_p}\rightarrow \varphi_{\m_1\dots\m_p}+a\,R_{\m_1\dots\m_p}$ can be rewritten in the form:
\begin{align}
\mathcal{L}=-\frac1{2(p+1)}\, (F_{\m_1\dots\m_{p+1}}+a\,Q_{\m_1\dots\m_{p+1}})\,(F^{\m_1\dots\m_{p+1}}+a\,Q^{\m_1\dots\m_{p+1}})\nn-\frac{1}{(p+1)\, (p+1)!}\e_{\m_1\dots\m_{p+1}\n_1\dots\n_{p+1}}\, a\,F^{\n_1\dots\n_{p+1}}\,Q^{\m_1\dots\m_{p+1}}\,,\label{ChiralpNew1}
\end{align}
where $Q_{\m_1\dots\m_{p+1}}=(p+1)\partial_{[\m_1}R_{\m_2\dots\m_{p+1}]}$.
The form of the action \eqref{ChiralpNew1} manifests two abelian gauge symmetries of the $p-$forms $\varphi_{\m_1\dots\m_p}$ and $R_{\m_1\dots\m_p}$.
At the same time, similarly to the original action \eqref{Chiralp}, fixing constant $a$ \eqref{ChiralpNew1} gives a single non-chiral $p-$form action. This discontinuity is tracked also in the fact that the PST form of the action \eqref{ChiralPST} is singular for the constant $a$ or other configurations with $c_\m\,c^\m=\partial_\m\,a\,\partial^\m\,a=0$.
Finally, one can rewrite the action in the form, resembling \eqref{ChiralpMin}:
\begin{align}
\mathcal{L}=-\frac1{2(p+1)}\, F_{\m_1\dots\m_{p+1}}\,F^{\m_1\dots\m_{p+1}}-\frac1{2(p+1)}\, (\cF_{\m_1\dots\m_{p+1}}+a\,Q_{\m_1\dots\m_{p+1}})\,(\cF^{\m_1\dots\m_{p+1}}+a\,Q^{\m_1\dots\m_{p+1}})\,,\label{ChiralpNew2}
\end{align}
%One non-standard feature of the action \eqref{Chiralp} is that it is not quadratic, despite the fact that it describes a free theory.
Combining the equations of motion $E^\varphi\,,\, E^R$ for the fields $\varphi_{\m_1\dots\m_{p}}$ and $R_{\m_1\dots\m_{p}}$ one gets\footnote{We use the notation: $$Q^\pm_{\m_1\dots\m_{p+1}}=Q_{\m_1\dots\m_{p+1}}\pm\frac{1}{(p+1)!}\e_{\m_1\dots\m_{p+1}\n_1\dots\n_{p+1}}Q^{\n_1\dots\n_{p+1}}$$ for (anti)self-dual part of the $(p+1)-$form $Q_{\m_1\dots\m_{p+1}}$.}
\begin{align}
E^R_{\m_2\dots \m_{p+1}}+a\,E^\varphi_{\m_2\dots\m_{p+1}}=\partial^{\m_1}a\, P_{\m_1\dots\m_{p+1}}=0\,,\\ P_{\m_1\dots\m_{p+1}}\equiv\cF_{\m_1\dots\m_{p+1}}+a\,Q^{+}_{\m_1\dots\m_{p+1}}\,,
\end{align}
which implies
\begin{align}
P_{\m_1\dots\m_{p+1}}=0\,,
\end{align}
automatically satisfying the equation of motion $E^a$ for the $a$ field,
\begin{align}
E^a=Q_{\m_1\dots\m_{p+1}}\,P^{\m_1\dots\m_{p+1}}=0\,.
\end{align}
This indicates the existence of a PST like symmetry shifting the scalar field $a$ which should have been expected given the equivalence to the PST formulation.

An interesting generalisation of the Lagrangian \eqref{ChiralpNew1} is (suppressing Lorentz indices):
\be
\mathcal{L}=-\frac12\, f(a)\,(\sqrt{a}\,F+\frac1{\sqrt{a}}\,Q)^2+\,f(a)\,F\wedge\,Q\,.\label{Lpm}
\ee
For $f(a)\sim 1/a$, this Lagrangian is equivalent to \eqref{ChiralpNew1} and describes a single chiral p-form carried in field $\vf$. For $f(a)\sim a$, it describes an anti-chiral $p-$form field carried by $R$.
The exchange $\vf \leftrightarrow R\,, a\rightarrow -\frac1{a}\,, f(a)\rightarrow -f(a)$ is a symmetry of the Lagrangian \eqref{Lpm}. %Interestingly, the latter condition stems from $a\to -a^{-1}$ in particular when $f(a)=a+a^{-1}$, i.e. when \eqref{Lpm} is a direct sum of chiral and anti-chiral actions with the same field variables.

%Another interesting observation is that the inversion of the ``background matrix'' in \eqref{Lpm} corresponds to $a\rightarrow -\frac1{a}, f(a)\rightarrow -\frac1{f(a)}$. 

%The interpretation of \eqref{rrL} as a duality-symmetric theory is not completely clear.

%Rewriting chiral cases of \eqref{Lpm} in terms of $\vf=\vf_++\vf_-\,,\; \tilde{\vf}=\vf_+-\vf_-$ for the special choices of $f(a)=a^{\pm 1}$ (with consecutive replacement $a\to a^{-1}$ in the second case) gives:
%\begin{align}
%\mathcal{L}_{\pm}=-\frac18 \, [(a+1)\,d \vf\pm (a-1)\,d \tilde{\vf}]^2+\frac14\,a\,d \vf\wedge\,d \tilde{\vf}\,,.\label{Lpm1}
%\end{align}
%where different signs correspond to different chiralities. The two actions transform into each other under $\vf\leftrightarrow \tilde{\vf}\,,\; a\rightarrow -a$.

The Lagrangian \eqref{ChiralpNew1} can be also rewritten as\footnote{$\star F$ is the Hodge dual of $F$ and $\wedge$ denotes the exterior product of forms.}:
\begin{align}
\cL=-\frac{p!}{2}\,\Big{(} \cM_{IJ}\,F^I\wedge \star F^J+\,\cK_{IJ}\,F^I\wedge F^J\Big{)}\,,\label{TDSp}
\end{align}
with
\begin{align}
 \cM_{IJ}=\begin{bmatrix}
    1 & a \\
    a & a^2
\end{bmatrix}\,,\quad \cK_{IJ}=\begin{bmatrix}
    0 & a \\
   -a & 0
\end{bmatrix}\,,\quad F^I=\begin{bmatrix}
    F \\ Q
\end{bmatrix}\,,
\end{align}
where $F^I$ is a two-vector with $p+1$-form components, $\cM$ is a two-by-two matrix of rank one, while the ``background matrix'' $\cE=\cM+\cK$ is invertible. The same action with the inverted background matrix $\cE^{-1}$ describes the same degrees of freedom, exchanging the roles of p-forms $\vf$ and $R$. The inversion of the background matrix is a particular $sl(2,R)$ rotation of the two-vector $F^I$ and therefore a field redefinition.
A potential generalisation of \eqref{TDSp} to $N$ chiral $p-$forms would be extending the matrices $\cM_{IJ}$ and $\cK_{IJ}$ to $2N\times 2N$ matrices, where $\cM_{IJ}$ has rank $N$. The Lagrangian \eqref{ChiralpNew1} is self-dual with respect to dualisation of both $\vf$ and $R$ fields. The replacement of the field $\vf$ with its magnetic dual $\tilde\vf$ via $F=d\vf=\star d\tilde\vf$ renders the same Lagrangian with $\vf$ replaced by $\tilde{\vf}$. Same is true for the replacement $Q=dR=\star d\tilde{R}$.

Next we will take a closer look at particular examples in different dimensions.

\section{Chiral two-form in six dimensions}

The first example we consider is the chiral two-form in six dimensions. For that we introduce two-forms $B_{\m\n}, R_{\m\n}$ and a vector field $c_\m$, as well as a Lagrange multiplier $G^{\m\n}$. The Lagrangian \eqref{Chiralp} is given in the following form:
\be
\mathcal{L}=-\frac16\, H_{\m\n\l}\,H^{\m\n\l}-\frac16\,(\mathcal{F}_{\m\n\l}-3\,c_{[\m}\,R_{\n\l]})\,(\mathcal{F}^{\m\n\l}-3\,c^{[\m}\,R^{\n\l]})+G^{\m\n}\partial_{[\m}c_{\n]}\,,\label{New6d}
\ee
where we denote:
\ba
\mathcal{F}_{\m\n\l}=H_{\m\n\l}+\frac16\,\epsilon_{\m\n\l\a\b\g}H^{\a\b\g}\,,\qquad H_{\m\n\l}=3\,\partial_{[\m}B_{\n\l]}\,.
\ea
The property of self-duality:
\be
\mathcal{F}_{\m\n\l}=\frac16\,\e_{\m\n\l\a\b\g}\mathcal{F}^{\a\b\g}\,,
\ee
implies:
\be
\mathcal{F}_{\m\n\l}\,\mathcal{F}^{\m\n\l}=0=\mathcal{F}_{[\m\n\l}\,\mathcal{F}_{\a\b\g]}\,.
\ee
Integrating out from the action \eqref{New6d} the auxiliary field $R_{\m\n}$ by solving its equations of motion,
\ba
\mathcal{F}_{\m\n\l}\,c^\l-c^2\,R_{\m\n}-2\,c_{[\m}R_{\n]\l}\,c^\l=0\quad \rightarrow\quad R_{\m\n}=\frac{1}{c^2}\mathcal{F}_{\m\n\l}\,c^\l+c_{[\m}\l_{\n]}\,,
\ea
one gets PST Lagrangian (up to total derivatives),
\be
\mathcal{L}=-\frac16\,H_{\m\n\l}\,H^{\m\n\l}+\frac{1}{2\,c^2}\,\mathcal{F}_{\m\n\l}\,c^\l\,\mathcal{F}^{\m\n\r}\,c_\r+G^{\m\n}\,\partial_{\m}\,c_\n\,,\label{PST6d}
\ee
which describes a free anti-self-dual two-form. The gauge symmetries of the Lagrangian \eqref{PST6d} can be extended to \eqref{New6d} by assigning appropriate gauge transformations for the new auxiliary field $R_{\m\n}$.
One of the symmetries of this system are given as \eqref{P6dsym}:
\ba
\d_{\a} B_{\m\n} = c_{[\m}\a_{\n]}\,,\quad \d_\a G^{\m\n}=\e^{\m\n\l\a\b\g}\a_\l\,c_\a\,R_{\b\g}\,,\quad \d_\a c_\m=0\,,\nonumber\\
\d_\a R_{\m\n}=-\partial_{[\m}\a_{\n]}+\frac{3}{2\,c^2}\P_-{}_{\m\n\l}{}^{\a\b\g}\,c^\l\,\partial_{\a}\,c_{\b}\,\a_{\g}\,.\label{6dsym}
\ea
It is straightforward to see, that the Lagrangian \eqref{New6d} and gauge transformations \eqref{6dsym} simplify for a choice of a background value for the pure gauge field $c_{\m}=\d_{\m}^6$. Nevertheless,  it is useful to keep manifest Lorentz symmetry as realised in \eqref{New6d}.
Another symmetry of the Lagrangian \eqref{New6d} is given as (see Appendix \ref{A} for details):
\ba
&\d_\varphi c_\m = \partial_\m \varphi\,,\quad \d_\varphi B_{\m\n}=\varphi\,R_{\m\n}\,,\quad
\d_\varphi G^{\m\n}=-\frac12\, \varphi\, \e^{\m\n\l\r\a\b}R_{\l\r}\,R_{\a\b}\,,
\nonumber\\
& \d_\varphi R_{\m\n}=\frac{3}{2\,c^2}\,\varphi\,c^\l\,\Pi_-{}_{\m\n\l}{}^{\a\b\g}\,\partial_{\a}R_{\b\g}\,.
\ea
Another symmetry of the action \eqref{New6d} is given by a transformation of $R_{\m\n}$ field only:
\be
\d_\phi R_{\m\n}=c_{[\m}\,\phi_{\n]}\,,
\ee
These symmetries are finite-step reducible. For example, choosing:
\be
\a_\m=c_{\m}\,\phi\,,\quad \phi_{\m}=\partial_\m \phi\,,
\ee
we get
\be
\d_\a B_{\m\n}=0\,,\quad \d_\a G^{\m\n}=0\,, \quad \d_{\a+\phi} R_{\m\n}\approx 0\,,
\ee
where the last identity holds on $G^{\m\n}$-shell.

Any consistent non-abelian extension of \eqref{New6d} is expected to possess same number of symmetries as the abelian action, deforming these symmetry transformations by terms proportional to coupling constants of non-abelian interactions.

%\pagebreak

\section{Chiral boson in two dimensions}

In two space-time dimensions of Minkowski signature, one can define the polynomial action for the free Chiral boson \eqref{Chiralp} in the following form:
\be
S_{New}=\int \Big(-\frac12 \partial_\m\vf\, \partial^\m \vf -\frac12 (\mathcal{F}^\m - c^\m\,R)(\mathcal{F}_\m - c_\m\, R) + \l\, \e^{\m\n}\partial_\m\, c_\n \Big)\, d^2\,x\,,\label{New2d}
\ee
where we introduce notations 
\be
\mathcal{F}_\m=\partial^+_\m \vf\,,\quad \partial^\pm_{\m}=\partial_\m \pm \e_{\m\n}\partial^\n\,.
\ee
The field $R$ is auxiliary, and can be integrated out, solving its algebraic equation, which gives:
\be
R=\frac1{c^2}c^\m\,\mathcal{F}_\m\,,
\ee
Plugging this back in the action, we get Pasti-Sorokin-Tonin form of the action:
\be
S_{PST}=\int \Big(-\frac12 \partial_\m \vf\,\partial^\m \vf+\frac{1}{2\,c^2}c^\m \, \mathcal{F}_\m\,c^\n \, \mathcal{F}_\n+\l\, \e^{\m\n}\partial_\m c_\n \Big)d^2\,x\,.\label{2dPST}
\ee
The latter action has the following gauge symmetry:
\be
\d c_\m=\partial_\m \a\,, \quad \d \vf = \a\,\frac1{c^2} c^\m\mathcal{F}_\m\,,\quad \d \l=\a\Big(\frac{c^\m \mathcal{F}_\m}{c^2}\Big)^2\,,\label{2dPSTSymmetry}
\ee
which can be shown by direct computation, using the identity $
\mathcal{F}_{\m}=\e_{\m\n}\mathcal{F}^\n\,.$

One can pull back the symmetry \eqref{2dPSTSymmetry} to the polynomial action \eqref{New2d} (as explained, e.g., in \cite{Henneaux:1992ig} (exercise 3.17)).
The difference between the actions \eqref{New2d} and \eqref{2dPST} is given as:
\be
S_{New}-S_{PST}=\int \Big(-\frac1{2\,c^2}(c^\m\,\mathcal{F}_\m-c^2\,R)^2\Big)\,d^2x \,,
\ee
and involves the square of the equations of motion for the field $R$, as expected. Therefore, the gauge transformation \eqref{2dPSTSymmetry} of the new action can be set to zero by assigning a gauge transformation rule for the field $R$. One can also use the equations of the auxiliary field to simplify gauge transformations.
For the action \eqref{New2d}, one can recast the symmetry as:
\be
\d c_\m=\partial_\m \a\,,\quad \d \vf=\a\, R\,,\quad \d \l=\a\, R^2\,,\quad \d R=\a\, \frac1{c^2} c^\m\partial^-_\m\,R\,.\label{GT2d}
\ee
The action \eqref{New2d} is polynomial, and among the gauge transformations \eqref{GT2d} only that of the new auxiliary field $R$ is non-polynomial.
In order to show that the action \eqref{New2d} is gauge invariant with respect to transformations \eqref{GT2d} it is helpful to make use of the identity:
\be
\mathcal{F}_\m=\frac1{c^2}(\e_{\m\n}c^\n+c_\m)\,c^\l\,\mathcal{F}_\l\,.
\ee
One can %redefine the fields in the action \eqref{New2d} as $b^{\m}=c^\m\,R\,,\,\, R\rightarrow \frac{1}{R}$, after which the auxiliary field $b_\m$ can be integrated out. The resulting action can be achieved also by directly 
integrate out the $c_\mu$ field in \eqref{New2d}, solving its equation of motion algebraically:
\be
c_\m=\frac{1}{R}\mathcal{F}_\m+\frac{1}{R^2}\e_{\m\n}\partial^\n\tilde r\,,
\ee
and plugging back into action (renaming $\frac{1}{R}\rightarrow r$) to get:
\be
S=\int \Big(-\frac12 \partial_\m\vf\, \partial^\m \vf-\frac12\, r^2\,\partial_\m \tilde r\,\partial^\m \tilde r-r\,\mathcal{F}^\m\,\partial_\m \tilde r \Big)\,d^2x\,.\label{2dPSTl}
\ee
It is now straightforward to see that integrating out $r$ gives PST action \eqref{2dPST}, with $c_\m=\partial_\m \tilde r$:
\be
S=\int \Big(-\frac12 \partial_\m\vf\, \partial^\m \vf+\frac1{2\,\partial_\m \tilde r\,\partial^\m\tilde r}\,(\mathcal{F}^\m\,\partial_\m \tilde r)^2\Big)\,d^2x
\ee
One can also arrive to this action by integrating out the Lagrange multiplier $\l$ in \eqref{2dPST}.

The action \eqref{2dPSTl} can be written in the \eqref{ChiralpNew1} form:
\be
\mathcal{S}=\int \Big(-\frac12 (\partial_\m\vf+r\,\partial_\m\tilde{r})^2+\e^{\m\n}\partial_\m\vf\,r\,\partial_\n\tilde{r}\Big)\,d^2x\,.\label{2dr}
\ee
Note, that the coefficient between the two terms in the action \eqref{2dr} is fixed. Changing the sign of the second term will change the chirality of the only excitation, while any other coefficient will result in a theory with both chiral and anti-chiral degrees of freedom. %It is therefore clear that the action \eqref{2dr} has a gauge symmetry, which is spoiled for other choices of the coefficient. %We will discuss this symmetry later.
%A simple generalisation would be to take a more generic degenerate metric.

%One may also think of generalisation of the action \eqref{2dSD} (and it's supersymmetric counterpart) where $r$ is a matrix. Such a generalisation could be suitable for T-duality symmetric description of String Theory. 

The Lagrangian of \eqref{2dr}  has a discreet symmetry:
\be
r\rightarrow \tilde{r}\,,\quad \tilde{r}\rightarrow r\,, \quad \vf\rightarrow -\vf-r\tilde{r}\,.
\ee
%A simple field redefinition, $\vf\rightarrow \vf-\frac12 r\tilde{r}$ and subsequent $r\rightarrow 2r$ renders the discreet symmetry more manifest in the redefined Lagrangian:
%\be
%\mathcal{L}=-\frac12 (\partial_\m \vf+ r\overleftrightarrow{\partial}_\m \tilde{r})^2-2\, \e^{\m\n}\vf\,\partial_\m r\,\partial_\n \tilde{r}\,,\label{2dxx}
%\ee
%where $r\overleftrightarrow{\partial}_\m \tilde{r}=r\partial_\m \tilde{r}-\tilde{r}\partial_\m r$ and the discrete symmetry is given by:
%\be
%r\rightarrow \tilde{r}\,,\quad \tilde{r}\rightarrow r\,, \quad \vf\rightarrow -\vf\,.
%\ee

%Another interesting symmetry of this action is given by following transformations:
%\ba
%\d \vf= g(r)+\tilde{g}(\tilde{r})\,,\quad \d r=\tilde{f}(\tilde{r})\,,\quad \d \tilde{r}=f(r)\,,
%\ea
%with the conditions:
%\ba
%\frac{\partial g}{\partial r}=-r \frac{\partial f}{\partial r}\,,\quad \frac{\partial \tilde{g}}{\partial \tilde{r}}=-\tilde{f}(\tilde{r})\,,
%\ea
%leaving independent two smooth functions of one field variable.
Further generalisations of chiral/duality-symmetric scalar field in two dimensions are discussed in Appendix \ref{2d}.

%We can interpret the fields $r,\tilde{r}$ as components of a $sp(2)$ vector $\xi^a$. In that case, we could rewrite the Lagrangian \eqref{2dxx} as:
%\be
%\mathcal{L}=-\frac12 (\partial_\m \vf+ \xi^a\partial_\m \xi_a)^2- \e^{\m\n}\vf\,\partial_\m \xi^a\,\partial_\n \xi_a\,,\quad \xi_a=\ve_{ab}\xi^b\,,
%\ee
%where $a,b=1,2$ are $sp(2)$ indices and $\ve_{ab}=-\ve_{ba}\,,\; \ve_{12}=1=\ve^{12}$.
%This form of the action has rigid shift symmetries for all three fields and $sp(2)$ rotations by a symmetric matrix $\l_{ab}=\l_{ba}$:
%\be
%\d \vf=\a+\b_a\,\xi^a+\l_{ab}\,\xi^a\,\xi^b\,,\quad \d \xi_a=\b_a+\l_{ab}\xi^b\,.
%\ee
%The discrete symmetry now reads as:
%\be
%\xi^a\leftrightarrow \xi_a\,,\quad \vf\rightarrow -\vf\,.
%\ee

%\pagebreak

\section{Duality-symmetric electromagnetism in four dimensions}
\label{Maxwell4d}

Similarly to the scalar in 2d and two-form in 6d, one can write a duality-symmetric action of polynomial form for Maxwell field in four dimensional Minkowski space. The latter case requires doubled field content, similarly to its PST equivalent \cite{Pasti:1995tn}. The polynomial Lagrangian for this case will be given as:
\be
\mathcal{L}=-\frac18\, F^a_{\m\n}\,F^{a\,\m\n}-\frac18\, (\mathcal{F}^a_{\m\n}-2\,c_{[\m}\,R^a_{\n]})(\mathcal{F}^{a\,\m\n}-2\,c^{[\m}\,R^{a\,\n]})+G^{\m\n}\,\partial_{[\m}c_{\n]}\,,\label{DSV}
\ee
where $a,b=1,2$, and
\ba
F^a_{\m\n}=\partial_{\m}A^a_{\n}-\partial_{\n}A^a_{\m}\,,\quad \mathcal{F}^a_{\m\n}=F^a_{\m\n}-\frac12 \e_{ab}\,\ve_{\m\n\l\r} F^{b\,\l\r}\,,\\
\e_{ab}=-\e_{ba}\,,\quad \e_{12}=1= \e^{12}\,,\quad 
\ve_{0123}=1=-\ve^{0123}\,.
\ea
The following identities hold (Einstein summation rule is assumed for both types of indices):
\be
 \mathcal{F}^a_{\m\n}\,\mathcal{F}^{a\,\m\n}=0\,,\quad 
\ve_{\m\n\l\r}\mathcal{F}^{a\,\l\r}=2\,\e^{ab}\,\mathcal{F}^b_{\m\n}\,,
\ee
Solving the algebraic equations of motion for $R^a_\m$, we get the PST action:
\be
\mathcal{L}_{PST}=-\frac18 F^a_{\m\n}\,F^{a\,\m\n}+\frac{1}{4\, c_\m c^\m}\,\mathcal{F}^a_{\m\n}\,c^\n\,\mathcal{F}^{a\,\m\r}\,c_\r+G^{\m\n}\partial_{[\m}c_{\n]}\,.
\ee
%A useful identity,
%\be
%0=5\,c^\r\,c_{[\r}\ve_{\m\n\l\s]}\,\mathcal{F}^{a\,\l\s}=2\,c^2\,\e^{ab}\mathcal{F}^b_{\m\n}+4\,c_{[\m}\e^{ab}\,\mathcal{F}^b_{\n]\r}\,c^\r+2\,\e_{\m\n\l\r}\,c^\l\,\mathcal{F}^{a\, \r\s}\,c_\s\,,
%\ee
%implies:
%\ba
%\mathcal{F}^a_{\m\n}=-\frac{2}{c^2}\P_-^{ab}{}_{\m\n}{}^{\l\r} c_\l\,\mathcal{F}^b_{\r\s}\,c^\s\,,\label{Fid}\quad \P_{\pm}^{ab}{}_{\m\n}{}^{\l\r}=\frac12 (\d^{ab}\,\d_{\m\n}^{\l\r}\pm \e^{ab}\,\ve_{\m\n}{}^{\l\r})\,,
%\ea
%which is instrumental in deriving the twisted self-duality equations from the equations of motion of \eqref{DSV}. Using the identity \eqref{Fid} one shows that $\cF^a_{\m\n}\,c^\n=0$ implies $\cF^a_{\m\n}=0$. 
The analogue of \eqref{ChiralpNew2} in this case would be the following Lagrangian
\begin{align}
\mathcal{L}=-\frac1{8}\, F_{\m\n}^a\,F^a{}^{\m\n}-\frac1{8}\, (\cF^a_{\m\n}+a\,Q^a_{\m\n})\,(\cF^a{}^{\m\n}+a\,Q^a{}^{\m\n})\,%\\
%=-\frac1{8}\, (F_{\m\n}^a+a\,Q^a_{\m\n})\,(F^a{}^{\m\n}+a\,Q^a{}^{\m\n})-\frac18\, \e_{ab}\,\e_{\m\n\l\r}\,a\,F^a{}^{\m\n}\,Q^b{}^{\l\r},
\label{DualitySymmetric2}
\end{align}
where $Q^a_{\m\n}=\partial_{\m}\,R^a_\n-\partial_\n\,R^a_\m$. This Lagrangian describes a single Maxwell field, using four vectors and a scalar. It can be also written in the form, similar to \eqref{TDSp}:
\begin{align}
\mathcal{L}=-\frac1{8}\,\cM_{IJ}\, F_{\m\n}^I\,F^J{}^{\m\n}-\frac1{16}\,\cK_{IJ} \e^{\m\n\a\b}\, F^I_{\m\n}\,F^J_{\a\b}\,,\label{TDS1}
\end{align}
where
\begin{align}
 \cM_{IJ}=\begin{bmatrix}
    1 & 0 & a & 0 \\
    0 & 1 & 0 & a \\
    a & 0 & a^2 & 0 \\
    0 & a & 0 & a^2
\end{bmatrix}\,,\quad
\cK_{IJ}=\begin{bmatrix}
    0 & 0 & 0 & a \\
    0 & 0 & -a & 0 \\
    0 & -a & 0 & 0 \\
    a & 0 & 0 & 0 \\
\end{bmatrix}\,,\quad F^I=\begin{bmatrix}
    F^1 \\ F^2 \\ Q^1 \\ Q^2
\end{bmatrix}\,,
\end{align}
Field redefinitions can lead to different matrices $\cM$ and $\cK$.
The detailed study of the rich symmetries of this Lagrangian will be conducted elsewhere. 

%\pagebreak

\section{Conclusions}

In this work we have constructed a polynomial action for free chiral $p-$forms with manifest Lorentz symmetry, finite number of auxiliary fields and consistent with general covariance, contrary to the folklore scepticism about the possibility of such a formulation. We showed that it reproduces the non-polynomial action formulation by Pasti, Sorokin and Tonin upon gauge fixing and integrating out an auxiliary field. An interesting feature of this formalism is that it is available only in Lorentzian signature. The covariant actions found here have rich structure of symmetries, which will be studied in detail elsewhere. Seemingly conventional form of the action \eqref{TDSp} encourages to study more systematically kinetic terms with non-invertible field-dependent bilinear forms. This may open a Pandora box of a large number of unexplored possibilities.

The actions presented here can be reduced to the non-covariant formulations of \cite{Henneaux:1988gg,Tseytlin:1990nb,Schwarz:1993vs} upon gauge-fixing and integrating out auxiliary fields. The latter statement is true already for the PST formulation, therefore the equivalence to PST formulation makes it evident.
The dualization properties of the actions presented here are somewhat similar to those of PST actions (see, e.g., \cite{Maznytsia:1998xw}), therefore there could be an alternative formulation of the polynomial actions \eqref{Chiralp} and \eqref{ChiralpNew2} where the last term with Lagrange multiplier is replaced by $\l\, \partial^{\m}\,c_\m$. Then, the divergence of $c_\m$ (instead of the curl) is constrained to be zero by the e.o.m. of the Lagrange multiplier, implying that $c_\m$ is a dual curvature of a $(d-2)$-form field, thus reproducing an alternative  action (covariantisation of Zwanziger-type action \cite{Zwanziger:1970hk} instead of the Schwarz-Sen one \cite{Schwarz:1993vs}) for the chiral $p-$form. 

The formulation we derive here adds an auxiliary field, another $p-$form, on top of the minimal PST formulation.
It can be therefore described as ``doubling $p-$forms to describe half a $p-$form''. The new formulation is related to PST by a variant of Hubbard-Stratonovich transformation, up to subtleties related to gauge fixing. As discussed after eq. \eqref{TDSp}, there are non-trivial duality symmetries relating the ``physical'' field $\varphi$ and ``auxiliary'' field $R$. Therefore, we have reasons to expect that this formulation is more than a polynomial rewriting of the PST action. In any case, the usefulness of this formalism will be tested by its ability to capture non-trivial interactions. This is a work in progress \cite{wip}.

The problem of interactions for chiral $p-$forms (and $p-$forms in general) has a long history
\cite{Freedman:1980us,Henneaux:1995ts,Perry:1996mk,Pasti:1997gx,Bandos:1997ui,Schwarz:1997mc,Aganagic:1997zq,Deser:1997mz,Medina:1997fn,Henneaux:1997ha,Bekaert:1998yp,Bekaert:1999dp,Bekaert:1999sq,Bekaert:2000qx,Bekaert:2001wa,Bandos:2014bva} with large body of negative results about their non-abelian interactions (see, however, \cite{Ho:2008nn,Pasti:2009xc,Ho:2011ni,Samtleben:2011eb,Chu:2012um,Samtleben:2012fb,Bandos:2013jva,Huang:2018hho,Buratti:2019cbm}). We hope, that the new formulation of the free theory presented here may help in the problem of interactions. It has two (arguable) advantages compared to the Sen's formulation \cite{Sen:2015nph,Sen:2019qit} --- general covariance and formulation in terms of gauge potentials.
Indeed, the gravitational coupling of the chiral $p-$forms are automatically consistent \cite{Bekaert:1998yp,Bekaert:2000rh}, if there are no additional degrees of freedom in the theory, like in \cite{Sen:2015nph,Sen:2019qit}. Therefore, the actions \eqref{Chiralp} and \eqref{ChiralpNew2} can be promoted to a generally covariant ones replacing the Minkowski metric with the dynamical metric of Einstein gravity.

One immediate question that can be asked is whether the formulation of the chiral bosons described here is advantageous compared to the PST formulation. 
The first challenge in this direction would be to formulate the non-linear DBI action for a single M5 brane \cite{Perry:1996mk,Pasti:1997gx,Bandos:1997ui,Schwarz:1997mc,Aganagic:1997zq,Ko:2013dka,Ko:2015zsy} in the variables of \eqref{Chiralp} or \eqref{ChiralpNew2}. Another interesting problem would be the BRST quantisation in the lines of \cite{Bekaert:2000rh}.
The actions studied here can be also used to write a Lorentz and generally covariant polynomial action for $d=10$ Type IIB Supergravity using finitely many auxiliary fields, extending the results of \cite{DallAgata:1997gnw,DallAgata:1998ahf}.
Some results on interacting theories generalising the free actions presented here will be reported in \cite{wip}.

\acknowledgments

The author thanks Arkady Tseytlin for stimulating discussions that motivated this work and multiple helpful suggestions.
Useful discussions with Zhirayr Avetisyan, Dmitry Bykov, Franz Ciceri, Chand Devchand, Oleg Evnin, Euihun Joung, Axel Kleinschmidt, Stefan Theisen and, especially, Cedric Troessaert are also gratefully acknowledged. The author thanks Dmitri Sorokin for his detailed feedback on the draft and suggestions.
This work was partially supported by Alexander von Humboldt Foundation. 
The author thanks Erwin Schr\"odinger Institute in Vienna for hospitality during the Workshop ``Higher spins and holography'' and Imperial College London for hospitality during several visits in the process of this work.

\appendix

\section{Abelian chiral two-forms in six dimensions}
\label{A}

We detail here some formulas related to the symmetries of the action \eqref{New6d}. Part of the gauge transformations for this action are given as abelian gauge transformations of the $B_{\m\n}$ and $G^{\m\n}$,
\be
\d_\xi B_{\m\n}=\partial_{\m}\xi_{\n}-\partial_\n \xi_\m\,,\quad \d_\xi G^{\m\n}=\partial_\r \xi^{\m\n\r}\,,
\ee
where $\xi^{\m\n\r}$ is an antisymmetric tensor parameter.
There is another gauge symmetry of PST action, that should have a counterpart here. Transformations with respect to this symmetry for $B_{\m\n}$, $c_\m$ and $G^{\m\n}$ fields takes the following form:
\be
\d_{\a} B_{\m\n} = c_{[\m}\a_{\n]}\,,\quad \d_\a G^{\m\n}=\e^{\m\n\l\a\b\g}\a_\l\,c_\a\,R_{\b\g}\,,\quad \d_\a c_\m=0\,,\label{6dPSTsym}
\ee
while for the new field $R_{\m\n}$, as we will see, it will take the following form:
\be
\d_\a R_{\m\n}=-\partial_{[\m}\a_{\n]}\,,\label{6dvarR}
\ee
up to some trivial transformations that vanish on-shell.

In order to check this gauge symmetry, we derive the field equations for each field separately:
\ba
\frac{\d \mathcal{L}}{\d B_{\m\n}}=\partial_{\l} H^{\m\n\l}-3\,\partial_\l\Big(c^{[\m}\,R^{\n\l]}-\frac16 \e^{\m\n\l\a\b\g}c_{\a}\,R_{\b\g}\Big)\,,\\
\frac{\d \mathcal{L}}{\d G^{\m\n}}=\partial_{[\m}c_{\n]}\,,\quad
\frac{\d \mathcal{L}}{\d R_{\m\n}}=\mathcal{F}^{\m\n\l}\,c_{\l}-c^2\,R^{\m\n}-2\,c^{[\m}\,R^{\n]\l}\,c_{\l}\,,\\
\frac{\d \mathcal{L}}{\d c_{\m}}=\mathcal{F}^{\m\n\l}\,R_{\n\l}-c^\m\,R^{\n\l}\,R_{\n\l}-2\,c^\n\,R^{\l\m}\,R_{\n\l}-\partial_{\r}G^{\r\m}\,.
\ea
Now, we compute variations with respect to symmetries \eqref{6dPSTsym}:
\ba
\frac{\d \mathcal{L}}{\d B_{\m\n}}\d_\a B_{\m\n}=-\partial_{[\m}\,c_\n\,\a_{\l]}\Big(\mathcal{F}^{\m\n\l}-6\,\P_-{}^{\m\n\l}_{\a\b\g}\,c^\a\,R^{\b\g}\Big)\nonumber\\
+c_{\m}\,\partial_\n\,\a_\l\Big(\mathcal{F}^{\m\n\l}-3\,c^{[\m}\,R^{\n\l]}\Big)\,,\label{6dvarB}
\ea
where we define projectors to (anti-)self-dual tensors:
\be
\P_{\pm}{}^{\m\n\l}{}_{\a\b\g}=\frac{1}{12}(\d^{\m\n\l}_{\a\b\g}\pm \e^{\m\n\l}{}_{\a\b\g})\,.
\ee
These projectors satisfy:
\ba
&\P_\pm{}^{\m\n\l}{}_{\a\b\g}=\P_{\mp}{}_{\a\b\g}{}^{\m\n\l}\,,\quad  \P_\pm{}^{\m\n\l}{}_{\a\b\g}\,\P_{\mp}{}^{\a\b\g}{}_{\r\s\t}=0\,,\\
&\P_\pm{}^{\m\n\l}{}_{\a\b\g}\,\P_{\pm}{}^{\a\b\g}{}_{\r\s\t}=\P_\pm{}^{\m\n\l}{}_{\r\s\t}\,,
\ea
Variation \eqref{6dvarR} with the $R_{\m\n}$-field is given as:
\be
\frac{\d \mathcal{L}}{\d R_{\m\n}}\d_\a R_{\m\n}=-c_{\m}\,\partial_{\n}\,\a_{\l}\Big(\mathcal{F}^{\m\n\l}-3\,c^{[\m}\,R^{\n\l]}\Big)\,,
\ee
and cancels the second term in \eqref{6dvarB} while the variation of the $G^{\m\n}$, that is supposed to compensate the first term of the rhs of \eqref{6dvarB}, is given as:
\be
\d_\a G^{\m\n}=\a_\l\,\Big(\mathcal{F}^{\m\n\l}-6\,\P_-{}^{\m\n\l}{}_{\a\b\g}\,c^\a\,R^{\b\g}\Big)\,,\label{6dvarG}
\ee
which differs from the expression in \eqref{6dPSTsym}. In order to show the equivalence of the two expressions, we make use of the following identity:
\ba
0=c^\r\,c_{[\r}\,\e_{\m\n\l\a\b\g]}\mathcal{F}^{\a\b\g}=6\,c^2\, \mathcal{F}_{\m\n\l}-18\,c^\r\,c_{[\m}\mathcal{F}_{\n\l]\r}-3\,\e_{\m\n\l\r\a\b}\,c^\r\,\mathcal{F}^{\a\b\g}\,c_\g \qquad\qquad
\nonumber\\ 
= 6 c^2\,\mathcal{F}_{\m\n\l}
-18\,c^2\,c_{[\m}\,R_{\n\l]}-3\,c^2\,\e_{\m\n\l\a\b\g}\,c^\a\,R^{\b\g} \qquad\qquad
\nonumber\\
-3\,\e_{\m\n\l\r\a\b}\,c^\r\,\Big(\mathcal{F}^{\a\b\g}-3\,c^{[\a}\,R^{\b\g]}\Big)\,c_\g
-18\,c_{[\m}\Big(\mathcal{F}_{\n\l]\r}-2\,c_{\n}\,R_{\l]\r}-R_{\n\l]}\,c_\r\Big)\,c^\r \qquad\qquad
\nonumber\\
=6\,c^2\Big(\mathcal{F}_{\m\n\l}-6\,\P_+{}_{\m\n\l}{}^{\a\b\g}\,c_{\a}\,R_{\b\g}\Big)-9\,\P_+{}_{\m\n\l}{}^{\a\b\g}\,c_\a\frac{\d \mathcal{L}}{\d R^{\b\g}} \qquad\qquad\nonumber\\
\approx 6\,c^2\Big(\mathcal{F}_{\m\n\l}-6\,\P_+{}_{\m\n\l}{}^{\a\b\g}\,c_{\a}\,R_{\b\g}\Big)\,,\qquad\qquad\label{6dFid}
\ea
where the last identity holds on-shell (up to equations of motion for $R_{\m\n}$).
Assuming that $c^2\neq 0$, we deduce the following identity:
\be
\mathcal{F}_{\m\n\l}\approx 6\,\P_+{}_{\m\n\l}{}^{\a\b\g}\,c_{\a}\,R_{\b\g}\,.
\ee
This identity allows to rewrite the variation \eqref{6dvarG} as:
\be
\d_\a G^{\m\n}\approx 6\, \a_\l\,\Big(\P_+-\P_-\Big)^{\m\n\l}{}_{\a\b\g}\,c^\a\,R^{\b\g}=\e^{\m\n\l\a\b\g}\,\a_\l\,c_\a\,R_{\b\g}\,,
\ee
which is exactly same as the variation \eqref{6dPSTsym} guessed from the PST symmetries.

Since the gauge transformation of the field $G^{\m\n}$ was changed up to equations of motion of $R_{\m\n}$, it induces a modification in the gauge transformations of the latter field. We compute this change for completeness here and get:
\ba
\d_{\a} B_{\m\n} = c_{[\m}\a_{\n]}\,,\quad \d_\a G^{\m\n}=\e^{\m\n\l\a\b\g}\a_\l\,c_\a\,R_{\b\g}\,,\quad \d_\a c_\m=0\,,\nonumber\\
\d_\a R_{\m\n}=-\partial_{[\m}\a_{\n]}+\frac{3}{2\,c^2}\P_-{}_{\m\n\l}{}^{\a\b\g}\,c^\l\,\partial_{\a}\,c_{\b}\,\a_{\g}\,.\label{P6dsym}
\ea
The last term is zero on $G^{\m\n}$-shell.
Alternative form of these gauge transformations are given as:
\ba
\d_{\a} B_{\m\n} = c_{[\m}\a_{\n]}\,,\quad
\d_\a c_\m=0\,,\quad \d_\a R_{\m\n}=-\partial_{[\m}\a_{\n]}\,,\nonumber\\
\d_\a G^{\m\n}=\e^{\m\n\l\a\b\g}\a_\l\,c_\a\,R_{\b\g}+\a_\l\,\Big(\mathcal{F}^{\m\n\l}-6\,\P_+{}^{\m\n\l}{}_{\a\b\g}\,c^\a\,R^{\b\g}\Big)\,.
\ea
In this case, the last term is zero on $R_{\m\n}$-shell as shown in \eqref{6dFid}. We will use the formulation \eqref{P6dsym} unless otherwise specified.

There is yet another gauge symmetry of the PST action, $\varphi$-symmetry:
\be
\d_\varphi c_\m = \partial_\m \varphi\,,\quad \d_\varphi B_{\m\n}=\varphi\,R_{\m\n}\,,
\ee
In order for the action \eqref{New6d} to be gauge invariant with respect to this symmetry, one has to assign gauge transformations:
\be
\d_\varphi G^{\m\n}=-\frac12\, \varphi\, \e^{\m\n\l\r\a\b}R_{\l\r}\,R_{\a\b}\,,\quad \d_\varphi R_{\m\n}=\frac{3}{2\,c^2}\,\varphi\,c^\l\,\Pi_-{}_{\m\n\l}{}^{\a\b\g}\,\partial_{\a}R_{\b\g}\,.
\ee
The latter symmetry of the action can be checked by direct computation.

\section{(Non-)chiral scalar in two dimensions}\label{2d}

Here we discuss some interesting observations related to chiral boson action \eqref{2dr} and its generalisations.
The Lagrangian \eqref{2dr} can be also written as:
\ba
\mathcal{L}=-\frac12 G_{AB} \partial_\m \vf^A\,\partial^\m \vf^B+\frac12 B_{AB}\,\e^{\m\n}\,\partial_\m \vf^A\,\partial_\n \vf^B\,,\label{Lagrangian2d}\\
\vf^A=(\vf, r, \tilde{r})\,,\quad \vf_A=\d_{AB}\vf^B\,,\quad A,B,C=1,2,3\,,
\ea
where the ``sigma model metric'' $G_{AB}$ has rank one and does not depend on $\vf_1\equiv \vf$:
\ba
G_{AB}(\vf)=\begin{bmatrix}
    1 & -\a\, \vf_3 & \tfrac{(1+\a)}2 \, \vf_2 \\
    -\a\, \vf_3 & \a^2\, \vf_3^2 &  - \tfrac{\a(1+\a)}2 \, \vf_2\,\vf_3\\\
    \tfrac{(1+\a)}2 \, \vf_2 & - \tfrac{\a(1+\a)}2 \, \vf_2\,\vf_3 & \tfrac{(1+\a)^2}4\,  \vf_2^2
\end{bmatrix}\,,\quad
B_{AB}= \begin{bmatrix}
    0 & -\g \vf_3 & \b \vf_2 \\
    \g \vf_3 & 0 &  -\d \vf_1 \\
    -\b \vf_2 & \d \vf_1 & 0
\end{bmatrix}\,,\qquad
\ea
and $\b+\g+\d=1$. The parameters $\a, \b, \g$ are arbitrary numbers: different choices of these numbers are related by field redefinition or boundary terms in the action.% In the action \eqref{2dxx} it is chosen $\a=1=\d,\,\b=\g=0$.

%One can also take a direct sum of the two Lagrangians of the type \eqref{2dr} --- self-dual and anti-self-dual --- as a candidate to the duality-symmetric parity-even theory. The corresponding Lagrangian will be given as follows:
%\ba
%\mathcal{L}=-\frac12 (\partial_\m\vf_++r_+\,\partial_\m\tilde{r}_+)^2-\e^{\m\n}\vf_+\,\partial_\m r_+\,\partial_\n\tilde{r}_+\nonumber\\
%-\frac12 (\partial_\m\vf_-+r_-\,\partial_\m\tilde{r}_-)^2+\e^{\m\n}\vf_-\,\partial_\m r_-\,\partial_\n\tilde{r}_-\,,
%\ea
%Overall, there are six fields involved --- one triplet $(\vf_\pm,r_\pm,\tilde{r}_\pm)$ for each chirality.

%\pagebreak

\subsection{A generalisation}

A generalisation of the action \eqref{2dr} is given as in \eqref{Lpm}:
\be
\mathcal{L}=-\frac12\, f(r)\,(\sqrt{r}\partial_\m\vf_++\frac1{\sqrt{r}}\partial_\m\vf_-)^2+\e^{\m\n}\,f(r)\,\partial_\m\vf_+\,\partial_\n\vf_-\,,\label{rrL}
\ee
For $f(r)=1/r$, this action is equivalent to \eqref{2dr} and describes a single chiral scalar carried in field $\vf_+$. Instead, for $f(r)=r$, it describes an anti-chiral scalar carried by $\vf_-$.
The replacement $\vf_+ \leftrightarrow \vf_-\,,r\rightarrow -\frac1{r}\,, f(r)\rightarrow -f(r)$ is a symmetry of \eqref{rrL}. %Interestingly, the latter condition stems from $r\to -r^{-1}$ in particular when $f(r)=r+r^{-1}$, i.e. when \eqref{rrL} is a direct sum of chiral and anti-chiral actions with the same field variables.
%Another interesting observation is that the inversion of the ``background matrix'' in \eqref{rrL} corresponds to $r\rightarrow -\frac1{r}, f(r)\rightarrow -\frac1{f(r)}$. 

In terms of $\vf=\vf_++\vf_-\,,\; \tilde{\vf}=\vf_+-\vf_-$ the special choices of $f(r)=r^{\pm 1}$ (replacing further $r\to r^{-1}$ in the second case) give:
\begin{align}
\mathcal{L}_{\pm}=-\frac18 \, [(r+1)\partial_\m \vf\pm (r-1)\partial_\m \tilde{\vf}]^2+\frac14\,\e^{\m\n}\, r\,\partial_\m\vf\,\partial_\n \tilde{\vf}\,,.\label{TDSA}
\end{align}
where different signs correspond to different chiralities. The two actions transform into each other under $\vf\leftrightarrow \tilde{\vf}\,,\; r\rightarrow -r$.

\subsection{A duality-symmetric formulation for 2d scalar}

The action \eqref{2dr} can be rewritten in a classically equivalent form:
\be
\mathcal{S}=\int \Big(-\frac12 (\partial_\m\vf-A_\m)^2+\frac12\e^{\m\n}\,\vf\,F_{\m\n}+B^{\m}(A_\m-r\partial_\m \tilde{\vf})\Big)\,d^2x\,,\label{2dA}
\ee
where $F_{\m\n}=\partial_\m\,A_\n-\partial_\n\,A_\m$ and $B$ is a Lagrange multiplier.

Dropping the last term in \eqref{2dA}, one can write a simpler (quadratic) action:
\be
\mathcal{S}=\int \Big(-\frac12 (\partial_\m\vf-A_\m)^2+\frac12\e^{\m\n}\,\vf\,F_{\m\n}\Big)d^2x\,,
\ee
which, after integrating out $A_\m$ is equivalent to free scalar field action.
We can double the field content of this action, in the following way:
\be
\mathcal{S}=\int \Big(-\frac14 (\partial_\m\vf_+-A_\m)^2-\frac14(\partial_\m\vf_-+A_\m)^2+\frac14\e^{\m\n}\,(\vf_++\vf_-)\,F_{\m\n}\Big)d^2x\,.\label{2dT}
\ee
This action is now gauge invariant with respect to $\d A_\m=\partial_\m a(x)\,,\; \d \vf_\pm = \pm a(x)$. After integrating out the field $A_\m$, we arrive to free scalar action, depending only on the field $\vf=\vf_++\vf_-$.
Instead of solving the equation of motion for $A_\m$, we could use the equations of motion of all fields to show, that on-shell $A_\m$ is pure gauge, therefore we can fix a gauge:
\be
0=A_\m=\frac12(\partial^+_\m\vf_++\partial^-_\m\vf_-)\,,
\ee
where the second equation is the equation of motion for the field $A_\m$ with $\partial^{\pm}_\m\equiv \partial_\m\pm\e_{\m\n}\partial^\n$. It follows from here, that in the (on-shell!) gauge $A_\m=0$, we have:
\be
\partial^+_\m\vf_+=0\,,\quad \partial^-_\m\vf_-=0\,,
\ee
hence the names of the fields.
The action \eqref{2dT} can be rewritten in terms of $\vf=\vf_++\vf_-$ and $\tilde{\vf}=\vf_+-\vf_-$ in the following form:
\be
\mathcal{L}=-\frac14(\partial_\m \vf)^2-\frac14(\partial_\m \tilde{\vf})^2+\frac12 A^{\m}(\partial_\m \tilde{\vf}-\e_{\m\n}\partial^\n \vf)-\frac14 A_{\m}A^{\m}\,,
\ee
Here, the gauge symmetry is given by:
\be
\d A_\m=\partial_\m a\,,\quad \d \tilde{\vf}=a\,.
\ee
On-shell, one can show that $A_\m$ is pure gauge, therefore one can choose a gauge $A_\m=0$, which, together with the e.o.m. for $A_\m$:
\be
A_\m=\partial_\m \tilde{\vf}-\e_{\m\n}\partial^\n \vf\,,
\ee
implies that there is only one d.o.f. propagating, as opposed to the same action without $A^2$ term, which would describe two degrees of freedom (see, e.g., \cite{Tseytlin:1990nb}). The on-shell equivalence does not mean one can plug back $A_\m=0$ to the action. Instead, we can fix, e.g., an off-shell gauge $\tilde{\vf}=0$ which is incompatible with $A_\m=0$.
%There is a discreet symmetry in this action: 
%\be
%\vf_\pm\rightarrow \vf_\mp\,,\quad A_\m\rightarrow -A_\m\,.
%\ee
The fields $\vf$ and $\tilde{\vf}$ do not enter the action in a completely symmetric form. In particular, there is a gauge symmetry that allows to gauge fix $\tilde{\vf}$ to zero, but not $\vf$. In fact, if one changes the sign in front of the $A^2$ term, the roles of the fields $\vf$ and $\tilde{\vf}$ change. This is not a field redefinition though and therefore is not a symmetry of the action. We can achieve the change of the sign in that term by a redefinition: $A_\m\rightarrow \e_{\m\n}A^\n$, but that results in simultaneous exchange of $\vf$ and $\tilde{\vf}$ in the mixed term, therefore leaving the roles of $\vf$ and $\tilde{\vf}$ unchanged. Even though the field $\tilde{\vf}$ is pure gauge, it is essential that its kinetic term has the right sign. If we take the opposite sign keeping gauge invariance of the action in tact (which is tantamount to changing also the sign of the $A^2$ term), the action becomes topological with zero degrees of freedom.

\bibliography{HSrefs}
\bibliographystyle{JHEP}

\end{document}